\documentclass[9pt,twocolumn,twoside]{pnas-new}

\templatetype{pnasresearcharticle} 

\usepackage{amsmath}
\usepackage{subfig}
\usepackage[toc,page]{appendix}
\usepackage{blindtext}
\usepackage{natbib}
\usepackage{siunitx}
\usepackage{relsize}
\usepackage{latexsym}
\usepackage{graphicx}
\usepackage{siunitx}
\usepackage{booktabs}
\usepackage{comment}
\usepackage{makecell}
\usepackage{relsize}
\usepackage{array,multirow,tabularx}
\usepackage{subcaption}
\usepackage{subfig}
\usepackage{lipsum}

\makeatletter
\AtBeginDocument{%
  \fancypagestyle{firststyle}{%
    \fancyhf{}
    \fancyfoot[R]{\footerfont PNAS\hspace{7pt}|\hspace{7pt}\textbf{\today}\hspace{7pt}|\hspace{7pt}vol. XXX\hspace{7pt}|\hspace{7pt}no. XX\hspace{7pt}|\hspace{7pt}\textbf{\thepage\textendash\pageref{LastPage}}}
    \fancyfoot[L]{\footerfont\@ifundefined{@doi}{}{\@doi}}
    \renewcommand{\headrulewidth}{0pt}%
  }%
}
\makeatother

\begin{document}

\title{Invariant fractocohesive length in thermally aged elastomers}

\author[a]{Aimane Najmeddine}
\author[b]{Santiago Marin}
\author[c]{Zhen Xu}
\author[c]{Connor Thompson}
\author[c,d]{Guoliang Liu}
\author[b,1]{Maryam Shakiba}

\affil[a]{Department of Civil and Environmental Engineering, Princeton University, USA}
\affil[c]{Department of Chemistry, Virginia Tech, USA}
\affil[d]{Macromolecules Innovation Institute, Virginia Tech, USA}
\affil[b]{Department of Aerospace Engineering Sciences, University of Colorado Boulder, USA}

\leadauthor{Najmeddine}


\significancestatement{Elastomers are widely used but often face severe degradation that compromises their functionality and limits long-term use. Temperature-induced aging alters the polymer network, producing stiffening, brittleness, and ultimately premature fracture. In this work, we show that, although both fracture toughness (measured in pre-cut samples) and work of fracture (from unnotched stress–strain curves) decrease during aging, their ratio remains constant. This invariance bridges bulk property evolution with fracture resistance, allowing toughness of thermally-aged elastomers to be inferred from simple tensile tests. To our knowledge, such a relationship has not been previously proposed and provides a new framework for predictive modeling of elastomer failure under thermo-oxidative conditions.}

\authorcontributions{A.N. proposed the hypothesis and devised the experiments; S.M, Z.X, and C.T carried out the experiments; M.S and G.L supervised the work; and A.N. and S.M. wrote the paper with input from all authors.}

\authordeclaration{The authors declare no competing interest.}

\correspondingauthor{\textsuperscript{1}To whom correspondence should be addressed. E-mail: an3801@princeton.edu}


\begin{abstract}
The fractocohesive length -- the ratio between fracture toughness and work-to-fracture -- provides a material-specific length scale that characterizes the size-dependent fracture behavior of pristine elastomers. However, its relevance to thermally aged materials, where both toughness and work of fracture degrade dramatically, remains unexplored. Here, we demonstrate that despite severe thermal embrittlement, the fractocohesive length remains invariant throughout thermal aging, independent of temperature or duration. We verify this invariance experimentally for two elastomer systems (Styrene Butadiene Rubber and Silicone Rubber) at multiple aging temperatures for aging times up to eight weeks. This finding bridges a critical gap in fracture mechanics of aged polymers: while the evolution of work-to-fracture can be predicted from well-established constitutive models that track network changes (crosslink density and chain scission), the evolution of fracture toughness has lacked predictive frameworks. The invariance of fractocohesive length enables direct calculation of fracture toughness at any aging state from the predicted work of fracture, eliminating the need for extensive fracture testing on \textit{aged} elastomers and providing a crucial missing link for computational fracture predictions in aged elastomeric components.
\end{abstract}

\dates{This manuscript was compiled on \today}
\doi{\url{www.pnas.org/cgi/doi/10.1073/pnas.XXXXXXXXXX}}

\maketitle

\thispagestyle{firststyle}

\ifthenelse{\boolean{shortarticle}}{\ifthenelse{\boolean{singlecolumn}}{\abscontentformatted}{\abscontent}}{}

\makeatletter
\fancyhead[LO,RO,LE,RE]{} 
\renewcommand{\headrulewidth}{0pt}
\makeatother

\firstpage[5]{4}

\begin{figure*}[bht!]
    \centering
    \subfloat[]{\includegraphics[width=0.75\textwidth]{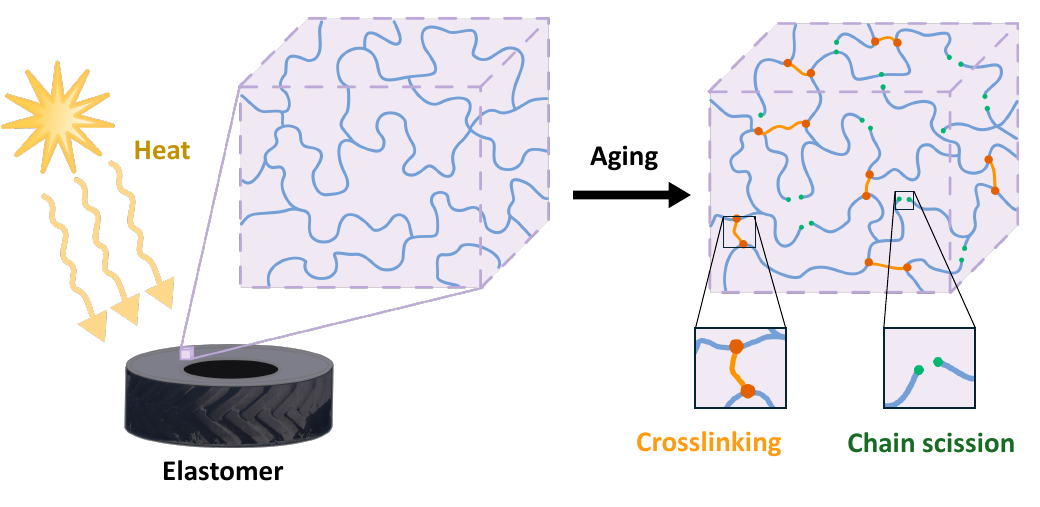}} \\
    
    \subfloat[]{\includegraphics[width=0.35\textwidth]{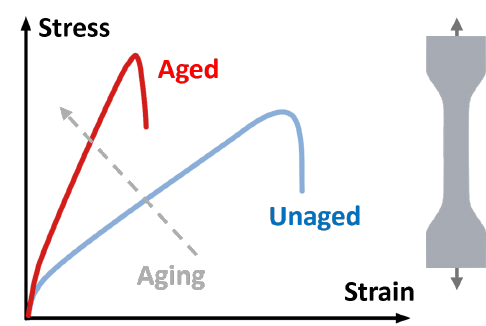}}

    \caption{(a) Schematic representation of the thermal aging process in elastomers. Exposure to elevated temperatures for extended durations can induce significant changes in the polymer network, including chain scission and crosslinking. 
    (b) After thermal aging, the mechanical response of elastomers can change, resulting in increased stiffness and reduced stretchability.}
    \label{fig1}
\end{figure*}

\dropcap{E}lastomers are widely used across industries due to their exceptional physical and mechanical properties. However, their extensive application exposes them to harsh conditions that compromise their functionality. Exposure to elevated temperatures, as illustrated in Figure~\ref{fig1}a for instance, triggers thermal degradation processes at the molecular level. The initially uniform elastomer network undergoes competing mechanisms of chain-scission and crosslinking, fundamentally altering its microstructure. Chain-scission breaks the elastomer backbone, reducing network connectivity, while crosslinking creates additional bonds between chains, increasing network density locally \citep{hamed1999tensile,CELINA2000171,SHAW20052758,celina2013review,johlitz2011chemical,JOHLITZ2014138}. These heterogeneous modifications result in an increase in the \textit{effective} crosslink density \cite{najmeddine2024physics}, producing a thermally aged brittle material with increased stiffness and reduced extensibility (Figure~\ref{fig1}b). 

While numerous studies have examined the effects of aging on elastomers' ultimate properties (strength and elongation at break) \cite{budrugeac1997accelerated,hamed1999tensile,gillen1997prediction,le1998methodologies,CELINA2000171,petrikova2011influence,JOHLITZ2014138,shakiba2021physics}, research on the fracture behavior of aged elastomers remains limited \cite{hernandez2019correlation,ha2008influence,anh2005effects,aglan2008effect,nichols1994effects}. Elastomeric fracture response can be characterized through two fundamental properties: (1) the work of fracture ($W_c$), measured as the area under the stress-strain curve in specimens without initial cuts, and (2) the fracture toughness or fracture energy ($G_c$), calculated as the energy required to advance a crack by unit area in pre-cut specimens. This gap severely limits computational predictions of failure in aged elastomers, as fracture simulations require toughness as an input parameter that currently must be experimentally determined for each aging condition.

The evolution of the work of fracture during thermal aging is well-documented. Changes induced by thermal aging can be represented through evolution functions of rubber modulus, chain extensibility, and failure stretch in physics-based hyperelastic constitutive frameworks \citep{rivlin1948large,ogden1972large,ArrudaBoyce93,gent1996new,ogden1997non}. These evolution functions can be derived from changes in the elastomer's macromolecular network properties during aging \cite{shakiba2021physics,najmeddine2024physics} or by solving thermo-chemical reaction kinetics occurring within the elastomer's network \cite{johlitz2011chemical,JOHLITZ2014138,CELINA2005395,Johlitz2013,KONICA2021104347,LION20121227,mohammadi2021constitutive,abdelaziz2019new}.

In contrast, the evolution of fracture toughness remains poorly understood. Specifically, there is no clear mathematical expression relating changes in toughness to the evolution of the elastomer's macromolecular network during thermal aging. The Lake-Thomas theory, which proposes that fracture energy scales with the product of chain density, dissociation energy of a single bond, and the square root of the number of bonds per chain, has proven unreliable for accurately estimating fracture toughness in practical polymer systems \cite{lake1967Strength}. This unreliability stems from two primary limitations: first, determining the dissociation energy of a single bond in a monomer unit within complex, heterogeneous network systems is exceedingly difficult; and second, the theory consistently underestimates experimental fracture toughness measurements -- often by an order of magnitude or more \cite{creton2016fracture}. This underestimation occurs because the Lake-Thomas model accounts only for chain scission at the crack tip while neglecting other significant energy dissipation mechanisms such as viscoelastic losses and molecular friction between chains that substantially contribute to measured fracture energy \cite{gent1996adhesion,long2015crack}. These limitations highlight the need for alternative approaches to understand fracture toughness evolution in aged elastomers. Thus, a critical need exists for a framework that connects the well-understood evolution of bulk properties during aging (i.e., work of fracture) to the complex evolution of fracture toughness.

The concept of a material-specific length-scale characterizing the size-dependent fracture behavior of pristine elastomers was first introduced by Thomas \citep{thomas1955rupture}, who suggested that fracture toughness scales with the work of fracture per unit volume multiplied by a length-scale variable representing the effective radius of the crack tip during tearing. Decades later, Chen et al. \cite{CHEN201750} revisited this proposal and confirmed that the ratio between fracture toughness (measured in the large-cut limit) and work of fracture (measured in the small-cut limit) defines a stable length-scale parameter for elastomers, which they termed the "fractocohesive length". This material-specific length corresponds to the critical crack length below which ultimate material properties remain independent of crack length \cite{CHEN201750,yang2019polyacrylamide}.

The existence of such a stable characteristic length offers a significant advantage: the ability to determine fracture toughness from work of fracture when their ratio is known. The fractocohesive length $\xi = G_c/W_c$ has been studied for unaged elastomers and gels \cite{song2021force,yang2019polyacrylamide,liu2019polyacrylamide,zheng2020fracture,yin2021essential,yin2021peel,long2021fracture,zhou2021flaw}, but no studies have examined this parameter for previously aged elastomers.

This study investigates the potential advantage of the fractocohesive length concept when analyzing the fracture response of aged elastomers. Here, we demonstrate a fundamental invariance: despite dramatic changes in both work of fracture and toughness during aging, their ratio -- the fractocohesive length -- remains constant, regardless of aging temperature or duration. This invariance provides the missing link between bulk property evolution and fracture behavior. This finding transforms how we numerically simulate fracture in aged elastomers: rather than requiring extensive fracture testing at each aging state, toughness can be predicted directly from simple tensile tests through constitutive models that track network evolution (e.g., effective crosslink density). We experimentally verify this hypothesis for two elastomer types under two different aging temperatures in the following sections. 

\section{Results and Discussion}

We validated the hypothesis that the fractocohesive length remains constant during thermal aging through an experimental investigation of two distinct elastomers: Styrene–Butadiene Rubber (SBR) and Silicone Rubber (SR). Samples were subjected to thermal aging at $\SI{70}{\celsius}$ and $\SI{110}{\celsius}$ for SBR, and at $\SI{120}{\celsius}$ for SR, for aging times ranging from $0$ to $8$ weeks. The selected temperatures allowed us to probe the hypothesis under both low and high thermal aging conditions. Sufficiently thin specimens (approximately $0.69~\si{mm}$ for SBR and $0.98~\si{mm}$ for SR) were prepared to ensure homogeneous aging, following the principles established by Celina et al.~\cite{celina2013review}.

\begin{figure*}[bht!]
    \centering
    \subfloat[]{\includegraphics[width=0.45\textwidth]{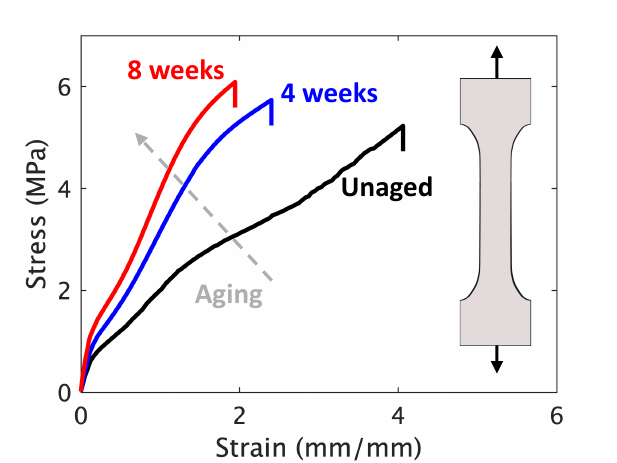}}
    \subfloat[]{\includegraphics[width=0.47\textwidth]{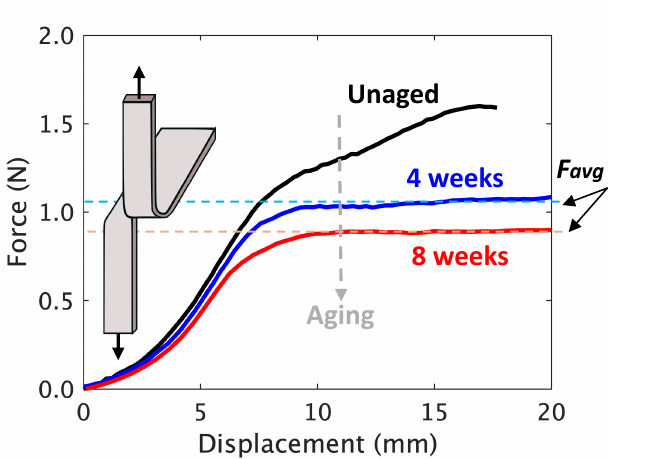}} \\
    
    \subfloat[]{\includegraphics[width=0.45\textwidth]{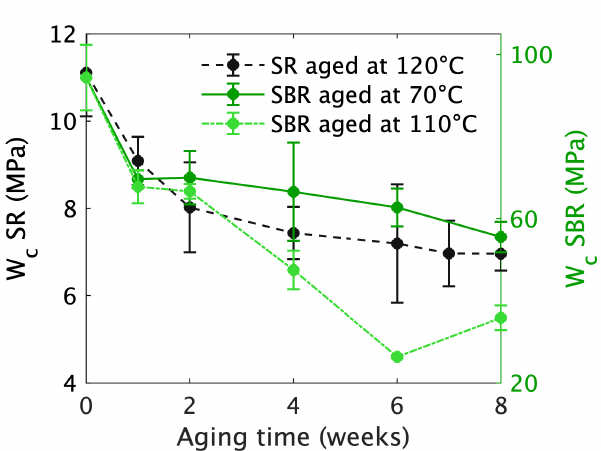}} \qquad
    \subfloat[]{\includegraphics[width=0.45\textwidth]{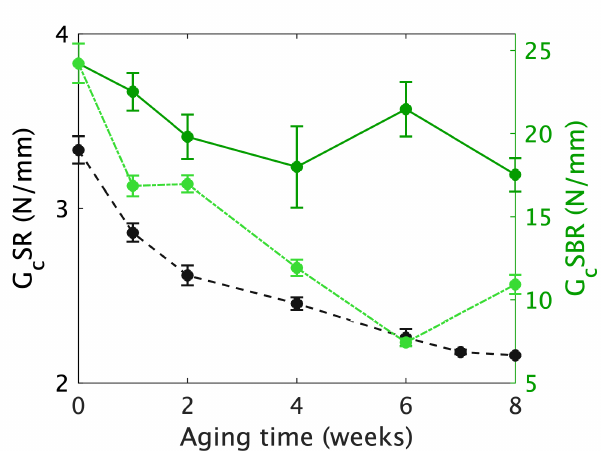}}

    \caption{(a) and (b) show the evolution of the stress–strain response during tensile tests and the force–displacement response during trouser tear tests, respectively, for SR after aging at $\SI{120}{\celsius}$. 
    (c) and (d) present the variation of the work of fracture, $W_c$, and the fracture toughness, $G_c$, respectively, with aging for SBR and SR.}
    \label{fig2}
\end{figure*}

Uniaxial tensile tests were performed on specimens without initial notches to measure the work of fracture. Representative stress–strain curves obtained at different aging times are shown in Figure~\ref{fig2}a. These results indicate that aging increases stiffness while reducing the failure strain of the elastomers, which in turn decreases the area under the stress–strain curve, i.e.,  $W_c$. Similarly, trouser tear tests were conducted on pre-cut specimens to measure the fracture toughness, $G_c$. Figure~\ref{fig2}b shows representative force–displacement curves, illustrating how the average force required for crack propagation decreases with aging time, leading to a reduction in $G_c$. Full sets of tensile and trouser-tear curves for both materials and all aging times are provided in the SI Appendix (Figures S1 to S3).

Figure~\ref{fig2}c and Figure~\ref{fig2}d present the evolution of $W_c$ and $G_c$ with thermal aging, respectively, for both materials and different aging conditions. Both properties show a substantial degradation over the aging period. Specifically, $W_c$ for SBR decreased from approximately 100 MPa to 60 MPa at $\SI{70}{\celsius}$ and to about 30 MPa at $\SI{110}{\celsius}$. Similarly, for SR,$W_c$ declined from about 10 MPa to 6–7 MPa during aging at $\SI{120}{\celsius}$. $G_c$ showed a comparable trend: for SBR, $G_c$ decreased from approximately 30 N/mm to 20 N/mm at $\SI{70}{\celsius}$ and to 10 N/mm at $\SI{110}{\celsius}$, while SR experienced a drop from 3.2 N/mm to 2.0 N/mm. These results underscore the substantial degradation in both bulk and localized mechanical properties induced by thermal aging.

\subsection{Invariance of Fractocohesive Length During Thermo-oxidative Aging}

\begin{figure*}[bht!]
    \centering
   \includegraphics[width=0.45\textwidth]{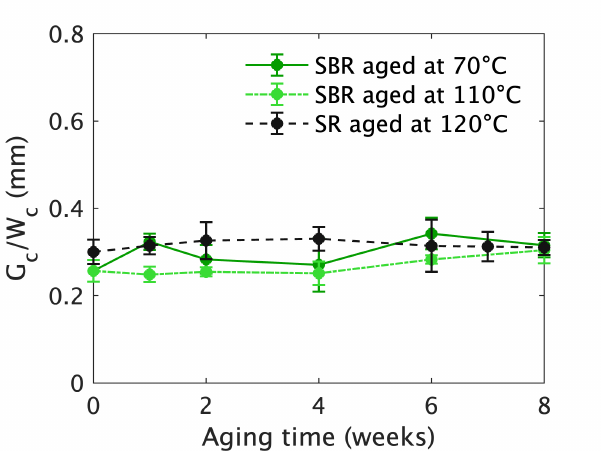} 
    \caption{Evolution of the fractocohesive length, $\xi = G_c / W_c$, for SBR and SR during aging at various exposure times and temperatures. The constant value of $\xi$ indicates that the fracture energy evolves in the same proportion as the work of fracture during thermal aging.}
    \label{fig:fig3}
\end{figure*}

Figure~\ref{fig:fig3} presents the evolution of the fractocohesive length, $\xi = G_c/W_c$, for SBR aged at $\SI{70}{\celsius}$ (dark continuous green line) and $\SI{110}{\celsius}$ (light dashed green line), and for SR aged at $\SI{120}{\celsius}$ (black dashed line). Despite the dramatic changes in the mechanical properties of both elastomers due to aging, the fractocohesive length remains invariant across all aging conditions and throughout the 8-week aging period. A weighted linear regression analysis confirmed no statistically significant dependence of $\xi$ on aging time for any condition, and a two-sample $t$-test indicated no significant difference between the mean values at $\SI{70}{\celsius}$ and $\SI{110}{\celsius}$. The slight fluctuations observed fall within the experimental error range, as indicated by the error bars, confirming the statistical validity of this invariance. This finding is particularly striking when contrasted with the substantial degradation in both $W_c$ and $G_c$ over the same period, as shown in Figure~\ref{fig2}c and Figure~\ref{fig2}d.

This parallel decline in both $W_c$ and $G_c$, maintaining their ratio constant, strongly supports our hypothesis that the fractocohesive length represents a fundamental material property that remains invariant during thermal aging. This suggests that while aging significantly alters the elastomer's underlying chain morphology, it affects both the bulk energy dissipation capacity and the localized fracture resistance proportionally. The weighted mean fractocohesive length, with standard error of the mean (SEM), was $0.303 \pm 0.011$ mm for SBR aged at $\SI{70}{\celsius}$, $0.266 \pm 0.006$ mm for SBR aged at $\SI{110}{\celsius}$, and $0.314 \pm 0.010$ mm for SR aged at $\SI{120}{\celsius}$. These values align well with the 0.1–1 mm range estimated from fracture surface roughness by Thomas \cite{thomas1960rupture} and Greensmith \cite{greensmith1960rupture}, as well as with more recent measurements on unaged elastomers by Chen et al. \cite{CHEN201750}.

The experimental confirmation of the fractocohesive length's invariance has a crucial implication: the material-specific length scale characterizing the size-dependent fracture behavior of elastomers (i.e., critical flaw size governing catastrophic failure) is not affected by thermal aging. To further explore this, we conducted a comprehensive flaw sensitivity analysis using phase-field fracture simulations. These simulations demonstrate that the transition from flaw-insensitive to flaw-sensitive failure occurs at the same critical cut depth for both unaged and aged elastomers, and this transition point corresponds precisely to the fractocohesive length determined experimentally.

\subsection{Phase-Field Modeling Framework for Flaw Sensitivity Analysis}

We employed a phase-field approach to model fracture propagation in single-edge notch specimens under uniaxial tension with varying initial cut depths. The phase-field method regularizes sharp crack discontinuities through a continuous damage field $d \in [0,1]$, where $d = 0$ represents undamaged material and $d = 1$ indicates complete failure \cite{miehe2010phase,najmeddine2024efficient,najmeddine2025coupled}. Appendix A summarizes the kinematics and governing equations pertaining to the coupled large-deformation phase-field solid mechanics problem solved in this work. Note that the AT1 version of the phase-field approach \cite{pham2011gradient} was used in this work. 

For the material constitutive response, we characterized the hyperelastic behavior of the elastomers using the Arruda-Boyce eight-chain model \cite{ArrudaBoyce93} to accurately incorporate network chain statistics that naturally evolve during thermal aging through changes in material properties, namely, rubber modulus and chain extensibility (i.e., limiting stretch).

A critical insight from the phase-field implementation concerns the relationship between the phase-field length scale $\ell$ and the fractocohesive length $\xi$. Through analytical derivation of the 1D phase-field solution of a 1D bar under uniaxial tension using the AT1 formulation, it was established that the phase-field length scale remains precisely $\ell = (3/16)\xi$ (See Derivations in the Appendix). 

Figure \ref{fig4}a demonstrates this relationship between the phase-field length scale (blue line) and the fractocohesive length (orange diamonds) -- through all aging times considered experimentally -- based on the AT1 analytical derivations. This relationship is fundamental: it demonstrates that the internal length scale governing damage localization in the phase-field model is directly proportional to the fractocohesive length, with a constant proportionality factor of 3/16. Importantly, this analytical result implies that if the fractocohesive length remains invariant during aging, so too must the phase-field length scale.

\begin{figure*}[bht!]
    \centering
    \subfloat[]{\includegraphics[width=0.45\textwidth]{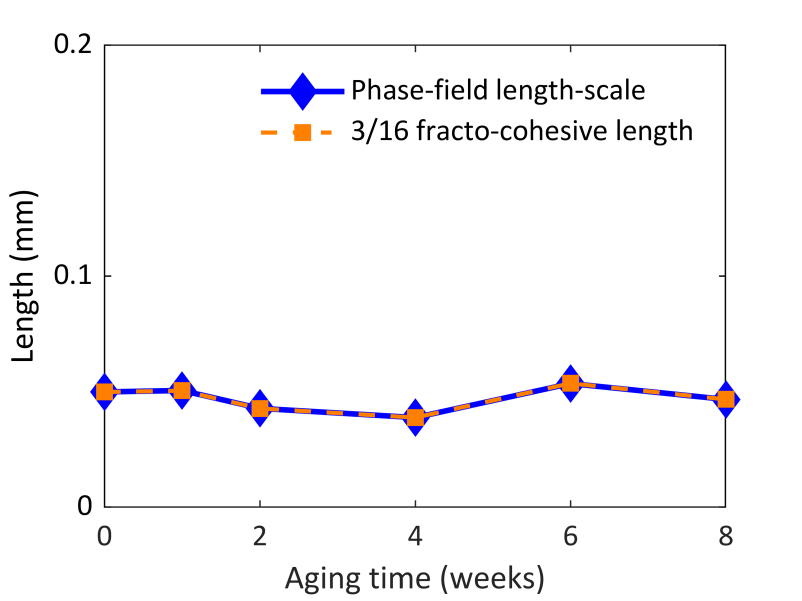}} \\
    
    \subfloat[]{\includegraphics[width=0.93\textwidth]{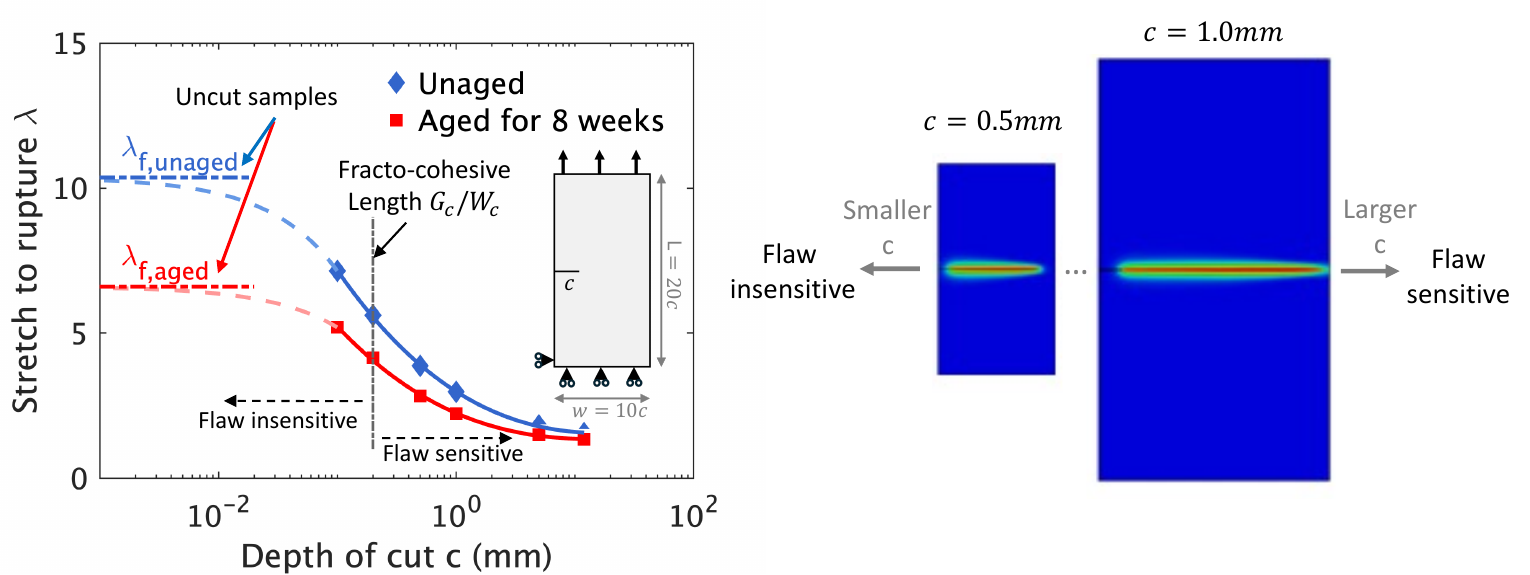}}

    \caption{(a) Relationship between phase-field length scale $\ell$ and fractocohesive length $\xi$ demonstrating the constant factor of 3/16 from AT1 formulation across all aging times. (b) Flaw sensitivity analysis showing stretch-to-rupture versus cut depth for unaged (blue) and aged (red) SBR, with the transition occurring at the same critical depth corresponding to the fractocohesive length (vertical gray dashed line)}
    \label{fig4}
\end{figure*}

The phase-field AT1-based relationship between the phase-field length-scale $\ell$ and the fractocohesive length $\xi$ aligns with recent work by Lee et al. \cite{lee2024size}, who postulated that the intrinsic length scale $G_c/W_c$ (equivalent to our fractocohesive length) can be directly incorporated into gradient-damage and phase-field models to capture size-dependent fracture. In their work, the phase-field length-scale was taken to be equal exactly to the intrinsic length scale that corresponds to the size of the fracture process zone, determined in their work precisely as the ratio $G_c/W_c$, i.e., the ratio between the macroscopic
critical energy release rate $G_C$ and the microscopic critical energy density $W_c$. 

To conduct 2D plane-stress uniaxial tension simulations of single-edge notched specimens, the finite element implementation was developed via a user-defined element (UEL) subroutine in ABAQUS for plane-stress conditions. To model aged materials, we modified only the bulk material parameters (shear modulus $\mu$, limiting stretch $\lambda_\text{limit}$, and fracture toughness $G_c$) based on our experimental measurements, while the phase-field length scale $\ell$ was obtained directly from $\ell = 3/16 \xi$. This approach tests whether the critical transition from flaw-insensitive to flaw-sensitive behavior, governed by the fractocohesive length, remains unchanged despite significant alterations in mechanical properties.

\subsection{Flaw Sensitivity Analysis Reveals Consistent Transition Behavior}

Figure \ref{fig4}b presents the stretch-to-rupture as a function of cut depth for both unaged SBR (blue diamonds) and SBR aged for 8 weeks at 70°C (red squares). Both conditions exhibit the characteristic transition from flaw-insensitive to flaw-sensitive rupture behavior, consistent with nonlinear elastic fracture mechanics predictions \cite{CHEN201750}.

In the flaw-insensitive regime (cut depths $c < 0.2~mm$), the stretch-to-rupture converges to the value obtained experimentally from uncut samples $\lambda_f$: approximately 10 for unaged samples and 6.5 for aged samples. As the cut depth increases beyond 0.2 mm, both configurations transition to a flaw-sensitive behavior, where the stretch-to-rupture decreases markedly with increasing cut depth. Notably, this transition occurs at the same cut depth for both materials — approximately 0.2 mm — which corresponds to the fractocohesive length, $\xi$ ($G_c/W_c$). The vertical gray dashed line marks this critical transition point.

\subsection{Implications of Invariant Length Scales}

The flaw sensitivity analysis provides compelling additional evidence for the invariance of fractocohesive length during thermal aging. While our primary experimental investigation examined two materials and aging temperatures and found consistent fractocohesive lengths across different aging times, the flaw sensitivity study demonstrates that this invariance is manifested in a critical fracture parameter: the threshold flaw size characterizing catastrophic fracture.

The analytical relationship $\ell = (3/16)\xi$ derived from the AT1 phase-field formulation provides theoretical grounding for this invariance. Since the phase-field length scale $\ell$ fundamentally controls damage localization and crack formation in the variational fracture framework, its proportionality to the fractocohesive length explains why both parameters remain constant during aging. This finding suggests that while thermal aging significantly alters the absolute values of mechanical properties, it preserves the fundamental length scales governing fracture processes.

\subsection{Physical rationale for an invariant fractocohesive length upon aging}

Both the bulk work of fracture in tension, \(W_c=\int_0^{\varepsilon_f}\sigma(\varepsilon)\,\mathrm{d}\varepsilon\), and the tearing/fracture energy, \(G_c=-\partial\mathcal{U}/\partial A\), are energetic functionals of the \emph{same} underlying network free-energy density \(W(\mathbf{F})\); they differ only in the loading geometry and the portion of the body over which the energy is accumulated. Foundational tear mechanics shows that \(G\) equals the change in \emph{stored} elastic energy when a cut is introduced or advanced in a stretched body, directly tying \(G_c\) to the bulk elastic energy scale that also sets \(W_c\) \citep{rivlin1953rupture,greensmith1963rupture,elmukashfi2021experimental}. At the molecular level, rubber elasticity links the small-strain shear modulus to the density of elastically active strands \(\mu\sim\nu_e k_BT\), where $nu_e$ is the chain density and $k_B$ is the Boltzmann constant, while finite-chain extensibility is governed by the strand length (molar mass between crosslinks \(M_c\)) or an equivalent finite-stretch parameter \(N\) in network models such as Arruda–Boyce \citep{Treloar1975,ArrudaBoyce1993}. The Lake–Thomas picture identifies the intrinsic part of \(G\) with the energy in highly stretched strands broken per unit crack area; it therefore scales with the number density and extensibility of load-bearing strands -- exactly the same chain-morphology controls that shape the tensile curve and hence \(W_c\) \cite{lake1967Strength,ahagon1975threshold}. By contrast, the \emph{length} of the near-tip process zone is governed not by an absolute energy but by \emph{ratios} of energetic measures—most notably the elasto-adhesive length \(\ell_{\mathrm{ea}}=\Gamma/E\)—so when a change in material state multiplies \(\Gamma\) and \(E\) by comparable factors, \(\ell_{\mathrm{ea}}\) (and the geometry of the active zone) remains nearly unchanged \citep{creton2016fracture,long2021fracture}. We use “mesoscale dissipation architecture” to mean the spatial organization, over \(\sim 1~\mu\mathrm{m}\)–\(1~\mathrm{mm}\), of the mechanisms that burn energy in the process zone—filler-network damage (Mullins), viscoelastic losses, micro-voiding/blunting, and, where applicable, strain-induced crystallization domains—whose characteristic span is set by \(\ell_{\mathrm{ea}}\) and related dissipative lengths.

Under \emph{homogeneous} thermal aging conditions (which is the case in this work), reaction kinetics proceed mainly via crosslink formation and chain scission; these reactions rescale \(\nu_e\) and \(M_c\) (hence \(\mu\) and \(N\)) throughout the body without introducing a new spatial pattern for the crack tip to "see" \citep{celina2013review}. Physically, homogeneous thermal aging acts like a uniform \emph{renormalization} of the network energy density: it co-rescales the bulk stress–strain curve \(\sigma(\varepsilon)\) and the near-tip traction–separation \(T(\delta)\) by nearly the same factors, while leaving the dissipation architecture (and its length scale) essentially unchanged. Because the same network knobs set both \(\sigma(\varepsilon)\) and \(T(\delta)\), the energy densities in bulk and at the crack tip co-evolve, so \(G_c\) and \(W_c\) change proportionally and the length-like ratio \(G_c/W_c\) remains approximately constant to leading order. This “energies rescale, length stays” view is consistent with modern soft-fracture frameworks where near-tip physics is governed by ratios of energetic measures such as the elasto-adhesive length \(\Gamma/E\) (a bulk elastic energy scale in the denominator), reinforcing that the same energy scale underlies both \(G_c\) and \(W_c\) \citep{creton2016fracture,long2021fracture}. 
Direct literature evidence supports the assertion that the \emph{same} chain-morphology controls co-govern tensile work-of-fracture and fracture energy. In gum NR/SBR/EPDM, systematic variations of crosslink \emph{density} and \emph{type} move tensile strength and tear strength along common trends (tear is usually more sensitive but responds to the same knob), demonstrating that both measures are driven by the same network parameters \citep{kok1986effects}. In filled EPDM, tensile properties and tear strength depend primarily on the overall crosslink density (network structure), again indicating common control variables for bulk and crack-tip energetics \citep{dijkhuis2009relationship}. Thermal-aging studies in carbon-black NR show that time/temperature exposure alters modulus, strength/elongation, and tearing strength \emph{together} (stiffening and reduced extensibility with a concomitant decline in tearing), i.e., \(W_c\) and \(G_c\) co-evolve under the same thermal aging kinetics \citep{li2015changes}. Chemical–mechanical correlations during oxidation in neoprene, and oxidation-driven modulus increases in polychloroprene, further tie crosslinking/scission kinetics to macroscopic mechanical energetics -- the very levers that set both \(W_c\) and \(G_c\) \citep{celina2000correlation,le2016predictive}. Comparable co-evolution has been reported in silicone systems, where aging increases modulus and ultimately degrades both tensile and tear resistance once scission dominates crosslinking, again pointing to shared network controls \citep{he2020thermo,mu2023study}. Taken together with the energetic identity linking tearing to stored bulk energy, these results justify anticipating an approximately constant \(G_c/W_c\) over aging time in thin, homogeneously aged specimens of fixed formulation; departures are expected only when aging \emph{restructures} the mesoscale dissipation architecture -- e.g., diffusion-limited oxidation (DLO) creating a brittle skin over a softer core, filler-network coarsening, or oxidation-induced porosity -- conditions under which the near-tip length scale itself changes \citep{celina2013review}.

\section{Conclusions}
We showed that the \emph{fractocohesive length} of elastomers, defined as \(\xi \equiv G_c/W_c\), remains essentially constant during homogeneous thermo-oxidative aging across the aging times and temperatures investigated. Experimentally, both the fracture energy \(G_c\) (trouser tear) and the bulk work of fracture \(W_c\) (area under the tensile curve) decrease with aging, but they do so in proportion, leaving \(\xi\) invariant within error. Physically, this invariance follows from the fact that uniform crosslinking/scission rescale the network’s elastic energy density and finite-extensibility \emph{uniformly}, affecting bulk and near-tip energetics in the same way, while the near-tip \emph{length} that organizes dissipation is set by mesoscale features (e.g., entanglements, filler spacing, etc) and by energetic ratios (e.g., \(\Gamma/E\)) that are not reorganized by homogeneous aging.

This co-scaling has practical consequences. First, it enables estimation of \(G_c(t,T)\) directly from tensile measurements via \(G_c=\xi\,W_c\), reducing experimental burden for aged materials. Second, because \(\xi\) is constant, the aging evolution of \(G_c\) can be expressed with a \emph{single} state variable tied to network morphology (e.g., effective crosslink density or an equivalent modulus/finite-stretch descriptor), which integrates cleanly with cohesive-zone and phase-field fracture models. In particular, combining (i) a kinetics-based evolution law for the network variable with (ii) the measured or modeled \(W_c(t,T)\) yields a self-contained predictor for \(G_c(t,T)\) that is immediately usable in numerical simulations of degraded elastomers.

The scope and limits are clear. The invariance of \(\xi\) is expected under \emph{homogeneous} aging—i.e., when oxidation is uniform and the near-tip dissipation architecture is not restructured. Departures should arise when diffusion-limited oxidation produces skin–core gradients, when filler networks coarsen or break down, or when other mechanisms qualitatively alter the process-zone microstructure or introduce new length scales (e.g., extreme rates or chemistry that suppresses/enhances strain-induced crystallization). These conditions provide targeted avenues for future work: deliberately inducing gradients or controlled architectural changes to quantify how \(\xi\) departs from constancy, and extending the present framework to multi-physics environments relevant to service aging. Overall, the demonstrated invariance of \(\xi\) consolidates a simple, physics-grounded bridge between bulk and fracture responses and offers a practical route to calibrate and forecast toughness degradation of elastomers under thermal oxidation.

\section{Materials and Methods} \label{sec: Experimental}

This section summarizes the experimental and modeling procedures used in this work. It first introduces the elastomeric materials selected for testing, then outlines the thermal aging protocols applied to them, and finally describes the mechanical tests used to quantify their fracture behavior. The section concludes with the theoretical and numerical framework employed to model damage and fracture using a variational phase-field formulation.

\subsection{Material} \label{sec: material}

Two elastomer systems have been identified to test the hypothesis that the fracto-cohesive length remains constant during thermo-oxidation: a filled Styrene–Butadiene Rubber (SBR) and Silicone Rubber (SR). Selection of the two rubber systems was made so as to investigate the validity of the hypothesis for two elastomers with noticeable difference in terms of their fracture behavior. SBR was acquired from Sumitomo Rubber Industries, Inc., while SR was purchased from McMaster-Carr. The composition of the filled SBR used is a mixture of butadiene and styrene-based polymer, which inherently has a higher tensile strength and stiffness. The SR used was composed of a silicone-based monomer, which inherently has a lower mechanical strength and stiffness. Each polymer is vastly different from the others in mechanical properties and chemical makeup. This was done in order to determine if the fracto-cohesive length would remain the same for carbon and silicone-based polymers.

\subsection{Thermal aging procedure} \label{sec: ThermoOxidationProcedure}

Thermo-oxidative aging was conducted in air-circulating convection ovens at controlled temperatures. SBR samples were aged at $\SI{70}{\celsius}$ and $\SI{110}{\celsius}$, while SR samples were aged at $\SI{120}{\celsius}$. Before aging, specimens were pre-cut into standardized geometries for uniaxial tensile and trouser tear tests, using a die cutter for the tensile specimens and a Cameo 3 plotter for the trouser tear specimens. The samples were suspended on metal racks to ensure uniform air exposure on all sides and were placed at the center of the preheated ovens. Specimens were aged for durations ranging from 0 to 8 weeks. For each aging period, a set of samples was removed from the oven, allowed to cool to room temperature, and subsequently subjected to mechanical testing.

\subsection{Mechanical testing} \label{sec: MechanicalTesting}

\subsubsection{Trouser tear test}

The fracture energy was determined using the trouser tear test, a classical method extensively described by Rivlin and Thomas~\citep{rivlin1953rupture}, Greensmith~\citep{greensmith1955rupture}, and Thomas~\citep{thomas1960rupture}. Rectangular specimens ($15~\si{mm}$ wide and $150~\si{mm}$ long) were prepared with a pre-cut slit forming two legs of $40~\si{mm}$ length. The legs were then pulled apart vertically in a universal testing machine (Figure~\ref{fig2}b).

To ensure a valid test configuration, the specimen thickness was kept small so that the bending moment in the legs was negligible compared to the tensile tearing force. Additionally, the leg lengths were made sufficiently long relative to the specimen width to promote tearing under simple extension conditions. Following established theoretical work~\citep{rivlin1953rupture,greensmith1955rupture,thomas1960rupture}, the fracture toughness $G_c$ was calculated as:
\[G_c = \frac{2F_{\mathrm{avg}}}{t}\]
where $t$ is the specimen thickness and $F_{\mathrm{avg}}$ is the steady-state tearing force, measured after the initial blunting stage when the force reached a plateau (Figure~\ref{fig2}b). Tests were performed at room temperature using a constant displacement rate of $50~\si{mm/min}$, in accordance with ASTM D624~\citep{standard2007standard}. At least five specimens were tested for each aging condition to ensure reproducibility and minimize experimental uncertainty.

\subsubsection{Tensile test}

Unnotched dogbone specimens were prepared according to the ASTM D638 standard. The narrow section measured $6~\si{mm}$ in width and $33~\si{mm}$ in length, with an overall specimen width of $19~\si{mm}$ and total length of $115~\si{mm}$ (Figure~\ref{fig2}a). The specimens were stretched to failure at a constant displacement rate of $50~\si{mm/min}$ at room temperature using an Instron universal testing machine. Due to the large deformations typical of elastomers, strain was measured by tracking the displacement of two markers placed on the narrow section using digital image analysis. The work of fracture, $W_c$, was obtained as the area under the engineering stress–strain curve. For each aging condition, at least five specimens were tested to improve statistical reliability and reduce experimental uncertainty.

\subsection{The variational phase-field theory and governing differential equations}

\subsubsection*{Kinematics}
A motion $\phi$ is defined as a one-to-one mapping $\mathbf{x} = \phi(\mathbf{X}, t)$ with a material point $\mathbf{X}$ in a fixed undeformed reference and $\mathbf{x}$ in a deformed spatial configuration with deformation gradient $\mathbf{F} \stackrel{\text{def}}{=} \frac{\partial \phi}{\partial \mathbf{X}}$. We define:
\begin{subequations}
\begin{align}
\bar{\mathbf{F}} &= J^{-1/3}\mathbf{F}, \quad J \stackrel{\text{def}}{=} \det \mathbf{F} \\
\mathbf{C} &= \mathbf{F}^T\mathbf{F} \\
\bar{\mathbf{C}} &= J^{-2/3}\mathbf{C}
\end{align}
\end{subequations}
where $\bar{\mathbf{F}}$ is the isochoric part of $\mathbf{F}$, $\mathbf{C}$ is the right Cauchy-Green tensor, and $\bar{\mathbf{C}}$ is the isochoric right Cauchy-Green tensor.

\subsubsection*{Free Energy and Damage}
The scalar damage field $d \in [0, 1]$ characterizes an intact state by $d = 0$ and a fully damaged state by $d = 1$. The free energy $\psi_R$ degraded by damage is:
\begin{multline}
\psi_R = \hat{\psi}_R(\mathbf{F}, \lambda_b, d, \nabla d) = g(d)\hat{\varepsilon}_R^0(\mathbf{F}, \lambda_b) \\
- \vartheta\hat{\eta}_R(\mathbf{F}, \lambda_b) + \hat{\psi}_{R,\text{nonlocal}}(\nabla d),
\label{eq:free_energy}
\end{multline}
with degradation function $g(d) = (1-d)^2$.

\paragraph{Nonlocal contribution}
\begin{equation}
\hat{\psi}_{R,\text{nonlocal}}(\nabla d)
= \frac{1}{2}\varepsilon_f^R\ell^2 |\nabla d|^2,
\label{eq:nonlocal_energy}
\end{equation}
where $\varepsilon_f^R$ is the bond dissociation energy per unit reference volume and $\ell$ is the intrinsic length scale.

\subsubsection*{Effective Bond Stretch}
The effective bond stretch $\lambda_b = L_t/L_0$ characterizes the bond-level elongation. The undamaged internal energy is:
\begin{equation}
\hat{\varepsilon}_R^0 = \frac{1}{2}Nn E_b(\lambda_b - 1)^2 + \frac{K}{2}(J-1)^2,
\label{eq:internal_energy}
\end{equation}
where $N$ is chain density, $n$ number of segments, $E_b$ bond stiffness, and $K$ the bulk modulus.

\paragraph{Configurational entropy (Arruda–Boyce)}
\begin{equation}
\hat{\eta}_R = -Nn k_b
\left[\frac{\bar{\lambda}\lambda_b^{-1}}{\sqrt{n}}
\left(\beta + \ln\left(\frac{\beta}{\sinh \beta}\right)\right)\right],
\label{eq:entropy}
\end{equation}
with $\beta = \mathcal{L}^{-1}\left(\frac{\bar{\lambda}\lambda_b^{-1}}{\sqrt{n}}\right)$ and $\bar{\lambda} = \sqrt{\frac{1}{3}\text{tr}\,\bar{\mathbf{C}}}$.

\subsubsection*{Stress Response}
The degraded first Piola–Kirchhoff stress is:
\begin{equation}
\mathbf{T}_R = 
\bar{\mu}\left(J^{-2/3}\mathbf{F} - \bar{\lambda}^2\mathbf{F}^{-T}\right)
+ (1-d)^2 K (J-1)J\mathbf{F}^{-T},
\label{eq:piola_stress}
\end{equation}
where
\[
\bar{\mu}
= \frac{Nk_b\vartheta}{3\sqrt{n}\bar{\lambda}\lambda_b}
\mathcal{L}^{-1}\left(\frac{\bar{\lambda}\lambda_b^{-1}}{\sqrt{n}}\right).
\]

\paragraph{Implicit bond stretch relation}

\subsection{1D Phase-Field Model for Uniaxial Tension with AT1 Formulation}

\subsubsection*{Problem Configuration}
We consider a one-dimensional bar of length $L$ subjected to uniaxial tension. The deformation is characterized by the stretch $\lambda = 1 + \frac{\partial u}{\partial X}$, where $u$ is the displacement field and $X$ is the material coordinate. For this analysis, we employ the AT1 (Ambrosio-Tortorelli Type 1) phase-field model, which introduces a damage threshold and exhibits different characteristics from the standard AT2 model.

\subsubsection*{AT1 Model Specifications}
The AT1 model is characterized by:
\begin{itemize}
    \item Geometric crack function: $\theta(d) = d$
    \item Derivative: $\theta'(d) = 1$
    \item Normalization constant: $c_0 = 4\int_0^1 \sqrt{\theta(d)} \, dd = 4\int_0^1 \sqrt{d} \, dd = \frac{8}{3}$
    \item Degradation function: $g(d) = (1-d)^2$
    \item Critical energy relation: $\varepsilon_f^R = \frac{G_c}{\ell}$
\end{itemize}

\subsubsection*{Kinematics and Constitutive Relations in 1D}

\paragraph{Deformation Measures.}
For a 1D bar under uniaxial tension:
\begin{align}
    x &= X + u(X,t) && \text{(current position)} \\
    F &= \frac{\partial x}{\partial X} = 1 + \frac{\partial u}{\partial X} = \lambda && \text{(deformation gradient)} \\
    J &= \lambda \lambda_2 \lambda_3 = 1 && \text{(incompressibility constraint)}
\end{align}
where the lateral stretches satisfy $\lambda_2 = \lambda_3 = \lambda^{-1/2}$ for incompressibility.

\subsubsection*{Governing Equations for the 1D Bar}

\paragraph{Damage Evolution Equation}
Following microforce balance, assuming steady-state conditions ($\dot{d} = 0$):
\begin{equation}
2(1-d)\mathcal{H} = \frac{G_c}{c_0}\left(\frac{\theta'(d)}{\ell} - 2\ell\nabla^2 d\right)
\label{eq:damage_general}
\end{equation}

For the AT1 model with $\theta'(d)=1$ and $c_0 = 8/3$:
\begin{equation}
2(1-d)\mathcal{H} = \frac{3G_c}{8}\left(\frac{1}{\ell} - 2\ell\nabla^2 d\right)
\label{eq:damage_at1}
\end{equation}

\paragraph{Uniform Bar Simplification}
For a uniform 1D bar where $\nabla^2 d = 0$:
\begin{equation}
2(1-d)\mathcal{H} = \frac{3G_c}{8\ell}
\label{eq:damage_uniform}
\end{equation}

Solving for the history variable:
\begin{equation}
\boxed{\mathcal{H} = \frac{3G_c}{16\ell(1-d)}}
\label{eq:history_damage}
\end{equation}

\subsubsection*{Damage Evolution Criterion}

\paragraph{Damage Initiation}
At damage onset ($d=0$):
\begin{equation}
\mathcal{H}_{\text{init}} = \frac{3G_c}{16\ell}
\label{eq:damage_initiation}
\end{equation}

Using the standard history field definition:
\begin{equation}
\psi_0 - \frac{G_c}{2\ell} = \frac{3G_c}{16\ell}
\end{equation}

Thus, the critical energy density for damage initiation is:
\begin{equation}
\boxed{\psi_0^{\text{crit}} = \frac{11G_c}{16\ell}}
\label{eq:critical_energy}
\end{equation}

\subsubsection*{Length Scale Relationship}

From equation \eqref{eq:history_damage}, we obtain:
\begin{equation}
\ell = \frac{3G_c}{16\mathcal{H}(1-d)}
\label{eq:length_scale}
\end{equation}

At damage initiation:
\begin{equation}
\boxed{\ell = \frac{3G_c}{16\mathcal{H}_{\text{init}}}}
\label{eq:length_scale_init}
\end{equation}

This demonstrates that the length scale is proportional to $G_c/\mathcal{H}_{\text{init}}$ with a prefactor of $3/16$, characteristic of the AT1 model.

\showmatmethods{} 

\acknow{This research was supported by the U.S. National Science Foundation (NSF) early CAREER award \#CMMI-2245155.}

\showacknow{} 

\bibsplit[32]

\bibliography{pnas-references}

\end{document}